\documentclass[aps,preprint,showpacs]{revtex4}%
\usepackage{amsfonts}
\usepackage{amsmath}
\usepackage{amssymb}
\usepackage{graphicx}%
\begin{document}
\title[ ]{ Geometry  and  off-shell  nilpotency  for $\mathcal{N}=1$  supersymmetric Yang-Mills theory }
\author{ A. Meziane\footnote { maptouk@yahoo.com; meziane.ahmed@univ-oran.dz}  and  M. Tahiri } \affiliation{Laboratoire de
Physique Th\'{e}orique d'Oran (LPTO), Universit\'{e} d'Oran, BP 1524
El M'Naouer, 31100 Es-Senia, Oran, Algeria}

\pacs{11.15.-q; 11.30.Pb;  12.60.Jv}

\begin{abstract}
 We show that for $\mathcal{N}=1$  supersymmetric Yang-Mills theory
it is possible to build an off-shell nilpotent BRST and anti-BRST
algebra in terms of a BRST superspace formalism.  This is based on
the introduction of  the basic fields of the quantized theory
together with an  auxiliary real field  via the lowest components of
the superfield components  of  a superYang-Mills connection. Here,
the associated supercurvature is constrained by horizontality
conditions as in ordinary Yang-Mills theory.  We also show  how the
off-shell BRST-invariant quantum action can be constructed starting
from a gauge-fixed superaction.

Keywords: Global supersymmetry; Auxiliary fields; Off-shell
nilpotent BRST and anti-BRST algebra; Quantum action.

\end{abstract}
\startpage{01}
\endpage{02}
\maketitle

\section{Introduction}

\bigskip
It is well known that the quantization of globally supersymmetric
gauge theories has been considerably studied a long time ago.
Several approaches have been proposed with various methods in order
to perform the quantization of such theories (for a review see e.g.
Ref. \cite{Hen} and references therein).

In the component field formalism, the supersymmetry algebra achieved
without auxiliary fields closes only on-shell. This can be explained
by the fact that the supersymmetry transformations of the models are
realized nonlinearly and therefore the main problem affecting such
theories is linked to its algebraic structure which involves
equations of motion and field dependent gauge transformations. This
gives rise to an infinite dimensional algebra, even if auxiliary
fields can be introduced to put the formalism off-shell
\cite{maison}. To avoid these difficulties it has been proposed in
Ref. \cite{magiorenico} the construction of a generalized BRST
operator by collecting together all the symmetries forming the
theory, namely ordinary BRST, supersymmetries and translations.
According to this procedure the role of the auxiliary fields is
covered by the external sources coupled to the nonlinear variations
of the quantum fields. This approach has been already successfully
applied to supersymmetric \cite{dixon1, white}, ordinary
\cite{magiore1} gauge field theories as well as to non gauge field
theories \cite{dixon2}.  Let us mention that in the Wess-Zumino
model it is only with auxiliary fields that one can obtain a tensor
calculus (for a review see e.g. Ref. \cite{Van-Nieuwen} and
references therein). .

Another possibility to solve the problem of the quantization of theories with on-shell algebra is to introduce the Batalin-Vilkovisky (BV) formalism \cite{bat22}. The
BV formalism is a very general covariant Lagrangian approach which
overcomes the need of closed classical algebra by a suitable
construction of BRST operator. The construction is realized by
introducing a set of the so-called antifields besides the fields
occurring in the theory. The elimination of these antifields at the
quantum level via a gauge-fixing procedure leads to the quantum
theory in which effective BRST transformations are nilpotent only
on-shell. Let us note, that the BV approach can be used to obtain
the on-shell BRST invariant gauge fixed action for $\mathcal{N}=1$
supersymmetric Yang-Mills theory in four dimensions without
requiring the set of auxiliary fields  \cite{Baulieu}.

Another interesting approach with infinite number of auxiliary
fields has been developed in the context of harmonic superspace
\cite{Galperin}. In this framework quantization of supersymmetric
theories has been discussed for various supersymmetries
\cite{Galperin2}.

On the other hand, it is also known that the extension of spacetime
with two ordinary anticommuting coordinates to a $(4,2$)-dimensional
superspace \cite{Bonora} leads in Yang-Mills type theories to
incorporate the gauge fields, the ghost and anti-ghost fields into a
natural gauge superconnection,  see also Ref.\cite{Malik} and
references therein. In such a superspace formalism the BRST and
anti-BRST transformations are derived systematically from the
horizontality conditions imposed on the supercurvature.

Let us note that the natural geometrical way to derive the BRST
structure of general gauge theories is to work, in the same spirit
as in Yang-Mills type theories, by using the superconnection
formalism. Within this framework and in contrast to what is done in
Yang-Mills theories, all superfield components of the supercurvature
cannot be constrained through horizontality conditions. This is a
consequence of the fact that the gauge theories we consider are
reducible and/ or open. It is the consistency of the Bianchi
identities  which is guaranteed by the remaining superfield
components of the supercurvature. Their lowest components allow the
introduction of auxiliary fields. These together with the fields
given by the lowest components of the superfield components of  the
superconnection represent the basic fields of the quantized gauge
theory. The off-shell nilpotency of the BRST and anti-BRST
transformations of these  fields is automatically ensured, thanks to
the structure equations and the Bianchi identities.  The BRST and
anti-BRST operators are related as usual to the partial derivatives
with respect to the two anti-commuting coordinates of the
superspace. Essentially, the introduction of the auxiliary fields
gives rise to the construction of the off-shell BRST invariant
quantum action. As shown in Ref. \cite{tahiri2} for the case of
non-Abelian BF theory  where the classical gauge algebra is
reducible and Ref. \cite{tahiri22} for the case of the simple
supergravity  where the classical gauge algebra is open, the
superspace formalism has been used in order to realize the BRST
structure of such theories. It leads to recast all the fields in
geometrical way and to introduce auxiliary fields ensuring the
off-shell invariance of the quantum action.

Our main aim in this paper consists to apply this formalism  for
discussing the off-shell nilpotent version of the BRST and anti-BRST
transformations for global $\mathcal{N}=1$, $4D$ supersymmetric
Yang-Mills theory where the classical gauge algebra is open
\cite{Van-Nieuwen,Wess}.  Let us mention that in Ref. \cite{tahiri2,
tahiri22}   the superspace formalism has been applied successfully
to theories with local symmetry while in the present work we are
interested to apply this formalism to a global supersymmetry. The
classical action for the $\mathcal{N}=1$ supersymmetric Yang-Mills
in four dimensions is given by \cite{Wess}
\begin{equation}
S_{0}=\int dx^{4} tr\large{(}- \frac{1}{4} F_{\mu\nu}F^{\mu\nu} -
\lambda\gamma^{\mu} D_{\mu}\overline{\lambda}\large{)},
\label{classical action1}%
\end{equation}
where $tr$ denotes the trace over the gauge algebra,
$(A_{\mu},\lambda)$ is the gauge multiplet, the field strenght is $
F_{\mu\nu}=
\partial_{\mu} A_{\nu}-\partial_{\nu} A_{\mu}
 + [ A_{\mu}, A_{\nu}]
 $  and  $D_{\mu}=\partial_{\mu} + [A_{\mu},]$
 is the covariant derivative.  By construction, in addition to the ordinary Yang-Mills symmetry, the action
(\ref{classical
 action1}) is invariant under the supersymmetry transformations
\begin{equation}
\delta_{\xi}A_{\mu}=
i\xi\gamma_{\mu}\overline{\lambda}  +   i\overline{\xi}\gamma_{\mu}\lambda,\nonumber\\
\end{equation}
\begin{equation}
\delta_{\xi}\lambda=\sigma^{\mu\nu}F_{\mu\nu}\xi,\label{transf1}
\end{equation}
 where $ \sigma^{\mu\nu}\equiv\frac{1}{2}[\gamma^{\mu},\gamma^{\nu}]$ and  $\xi$ is a spin $1/2$ valued
 infinitesimal supersymmetry parameter.

In the following we shall call the superspace obtained by enlarging
spacetime with two ordinary anticommuting coordinates BRST
superspace, in order to distinguish it from superspace of
supersymmetric theories. Let us recall that a full off-shell
structure of any supersymmetric field theory most naturally exhibits
itself in superspace, provided the superfield formulation of the
theory in terms of unconstrained superfields is available. The BRST
superspace formalism we present here permits us to derive the
off-shell nilpotent BRST and anti-BRST algebra of quantized
$\mathcal{N}=1$, $4D$ supersymmetric Yang-Mills theory.  In
particular, it gives another possibility leading to the minimal set
of auxiliary fields in such theory.

Our paper is organized as follows: In Section II, the BRST
superspace approach and horizontality conditions for
$\mathcal{N}=1$, $4D$ supersymmetric Yang-Mills theory are
discussed. We also show how the various fields of such theory and
their off-shell nilpotent BRST and anti-BRST transformations can be
determined via a BRST superspace formalism. The construction of the
BRST-invariant quantum action for $\mathcal{N}=1$ super Yang-Mills
theory in terms of this off-shell structure is described in Section
III.   Section IV is devoted to concluding remarks.

\section{Off-shell nilpotent BRST Algebra}
 \bigskip Let $\Phi$  be a super Yang-Mills connection on a $(4,2)$-dimensional BRST superspace with coordinates
 $z$=$(z^M)$=$(x^\mu,\theta^\alpha)$, where
 $(x^\mu)_{\mu_{_={1,..,4}}}$ are the coordinates of the spacetime
manifold and $(\theta^\alpha)_{\alpha_{_{=1,2}}}$  are ordinary
anticommuting variables. Acting the exterior covariant
superdifferential $D$ on $\Phi$ we obtain the supercurvature
$\Omega$ satisfying the structure equation,  $\Omega$ =
d$\Phi$+(1/2)[$\Phi$,$\Phi$],  and the Bianchi identity,
d$\Omega$+[$\Phi$,$\Omega$]= 0.  The superconnection $\Phi$ as
1-superform on the BRST superspace can be written as
\begin{equation}
\Phi=dz^{M}(\Phi^{i}_{M}I_{i}+\Phi^{\mu}_{M}P_{\mu} +
\Phi^{a}_{M}Q_{a}),
\end{equation}\label{phi}%
where $\{I_{i}\}_{i=1,..,  d = dim G}$ and $\{P_{\mu},Q_{a}\}_{\mu
=1,...4;   a = 1,..,4}$  are the generators of the internal symmetry
group $(G)$ and the $\mathcal{N}=1$  supersymmetric group $(SG)$
respectively. They satisfy the following commutation relations
\begin{equation}
[I_{i},I_{j}]=f^{k}_{ij} I_{k},\nonumber\\
\end{equation}
\begin{equation}
[I_{i},P_{\mu}]=[Q_{a},P_{\mu}]= [P_{\mu},P_{\nu}]=0,\nonumber\\
\end{equation}
\begin{equation}
[Q_{a},Q_{b}]=2(\gamma^{\mu})_{ab}P_{\mu},\nonumber\\
\end{equation}
\begin{equation}
[I_{i},Q_{a}]=b^{*}_{i}Q_{a},
\end{equation}\label{commutateur}%
where $\{\gamma^{\mu}\}_{\mu =1,..,4}$  are the Dirac matrices in
the Weyl basis, $b^{*}_{i}=b_{i}$ for $a =1,2$ and
$b^{*}_{i}=-b_{i}$ for $a=3,4$ giving the representation of the
internal symmetry of $Q_{a}$ and $[,]$ the graded Lie bracket. Let
us mention that the supersymmetric generators $\{Q_{a}\}$ are given
in the Majorana representation \cite{Wess,WEinberg}.  Note that the
Grassmann degrees of the superfield components of $\Phi$ are given
by $|\Phi^{i}_{M}|$=$|\Phi^{\mu}_{M}|=m$, $|\Phi^{a}_{M}|=m+1$ (mod
$ 2$), where $m=|z^{M}|$ $(m=0 $ for $M=\mu$ and $m=1$ for
   $M=\alpha$), since $\Phi$ is an even 1-superform.

However, we assign to the anticommuting coordinates $\theta^{1}$ and
$\theta^{2}$ the ghost numbers $(-1)$ and  $(+1)$ respectively, and
ghost number zero for an even quantity: either a coordinate, a
superform or a generator. These rules permit us to determine the
ghost numbers of the superfields ( $\Phi^{i}_{\mu}$,
$\Phi^{\nu}_{\mu}$, $\Phi^{a}_{\mu}$, $\Phi^{i}_{1}$,
$\Phi^{i}_{2}$, $\Phi^{\nu}_{1}$, $\Phi^{\nu}_{2}$, $\Phi^{a}_{1}$,
$\Phi^{a}_{2}$) which are given by $(0,0,0,+1,-1,+1,-1,+1,-1)$

    Upon expressing the supercurvature  $\Omega$ as
\begin{equation}
   \Omega=\frac{1}{2} dz^{N}\wedge dz^{M} \Omega_{MN} = \frac{1}{2}dz^{N}\wedge dz^{M}
   (\Omega^{i}_{MN}I_{i}+\Omega^{\mu}_{MN}P_{\mu}+\Omega^{a}_{MN}Q_{a}),\label{Supercurvature}
   \end{equation}
we find from the structure equation
\begin{subequations}
\label{allequations} 
\begin{eqnarray}
\Omega_{\mu\nu}&=&\partial_{\mu} \Phi_{\nu} - \partial_{\nu}
\Phi_{\mu} + [ \Phi_{\mu}, \Phi_{\nu}],\label{equationa}
\\
\Omega_{\mu\alpha}&=&\partial_{\mu} \Phi_{\alpha} -
\partial_{\alpha} \Phi_{\mu} + [ \Phi_{\mu}, \Phi_{\alpha}],\label{equationb}
\\
\Omega_{\alpha\beta}&=&\partial_{\alpha} \Phi_{\beta} +
\partial_{\beta} \Phi_{\alpha} + [ \Phi_{\alpha}, \Phi_{\beta}].\label{equationc}
\end{eqnarray}
\end{subequations}

Similarly, the Bianchi identity becomes

\begin{subequations}
\label{allequations} 
\begin{eqnarray}
D_{\mu}\Omega_{\nu\kappa} +
D_{\kappa}\Omega_{\mu\nu}+D_{\nu}\Omega_{\kappa\mu}&=&0,\label{equationa1}
\\
D_{\alpha}\Omega_{\mu\nu}-D_{\nu}\Omega_{\mu\alpha}+D_{\mu}\Omega_{\nu\alpha}&=&0,\label{equationb2}
\\
D_{\alpha}\Omega_{\beta\gamma}+D_{\beta}\Omega_{\alpha\gamma}+D_{\gamma}\Omega_{\alpha\beta}&=&0,\label{equationc3}
\\
D_{\mu}\Omega_{\alpha\beta}-D_{\alpha}\Omega_{\mu\beta}-D_{\beta}\Omega_{\mu\alpha}&=&0,\label{equationd4}
\end{eqnarray}
\end{subequations}
where $D_{M}=\partial_{M}+ [ \Phi_{M}, .]$ is the $M$ covariant
superderivative.  Now, we shall search for the constraints to the
supercurvature $\Omega$ in which the consistency with the Bianchi
identities  $(7)$ is ensured. This requirement ensures then the
off-shell nilpotency of the BRST and anti-BRST algebra. The full set
of supercurvature constraints turns out to be given by

\begin{equation}
\Omega_{\mu\alpha}=0,\quad\quad\quad\quad\quad\quad
\Omega_{\alpha\beta}=0.\label{constraint1}
\end{equation}
It is easy to check the consistency of this set of supercurvature
constraints through an analysis of the Bianchi identities. Indeed,
we remark that identities (\ref{equationc3}) and (\ref{equationd4})
are automatically satisfied because of the constraints
 (\ref{constraint1}) while identities (\ref{equationa1}) and
 (\ref{equationb2}) yield further restrictions on supercurvature
$\Omega$
\begin{equation}
\Omega^{\kappa}_{\mu\nu}=0 ,\quad\quad\quad\quad\quad\quad
 \Omega^{a}_{\mu\nu}=0. \label{constraints2}
\end{equation}
At this point, let us mention that the consistency of the
horizontability conditions (\ref{constraint1}) and
(\ref{constraints2}) with the Bianchi identities $(7)$, as we will
see later, guarantees automatically the off-shell nilpotency of the
BRST and anti-BRST transformations on all the fields belonging to
$\mathcal{N}=1$ super Yang-Mills theory.

 \bigskip Now in order to derive the off-shell BRST structure of $\mathcal{N}=1$ super
Yang-Mills theory using the above BRST superspace formalism, it is
necessary to give the geometrical interpretation of the fields
occurring in the quantization of such theory. Besides the gauge
potential $\Phi_{\mu |}^{i}=A_{\mu}^{i}$, there exists the
superpartener $\Phi_{a|}^{i}=\lambda_{a}^{i} $  of  $A_{\mu}$ and an
auxiliary real field $\Phi_{|}^{i}=\Lambda^{i}$ which are introduced
via the field redefinitions $\Phi^{ai}=
-\frac{1}{4}(\gamma^{\mu})^{ab}[Q_{b},\Phi_{\mu}^{i}]$ and
$\Phi^{i}=\frac{1}{4}\delta_{b}^{a}[Q_{a},\Phi^{bi}]$ respectively.
The components $\Phi_{\mu|}^{\nu} $  and $\Phi_{\mu|}^{a}$  are
identified to zero. Note that the symbol  $  " |"$ indicates that
the superfield is evaluated at $\theta^{\alpha}=0$.

Furthermore, we introduce the following now:
$\Phi_{1|}^{i}=c_{1}^{i}$ is the ghost for Yang-Mills
    symmetry, $\Phi_{2|}^{i}=\overline{c}_{2}^{i}$  is the antighost of  $c_{1}^{i}$, $B^{i}=\partial_{1}\Phi_{2|}^{i}$
    is the associated auxiliary field, $\Phi_{1|}^{a}=\chi_{1}^{a}$ is the supersymmetric
ghost,  $\Phi_{2|}^{a}=\overline{\chi}_ {2}^{a}$,
     is the antighost of $\chi_{1}^{a}$, $G^{a}=\partial_{1}\Phi_{2|}^{a}$  is the associated auxiliary field,
     $ \Phi_{1|}^{\mu}=\xi_{1}^{\mu}$ is the translation symmetry ghost, $\Phi_{2|}^{\mu}=\overline{\xi}_{2}^{\mu}$,
      is the antighost of $\xi_{1}^{\mu}$ and $E^{\mu}=\partial_{1}\Phi_{2|}^{\mu} $ is the associated
     auxiliary field. Let us mention that the symmetry ghosts and
      antighosts $\chi_{\alpha}^{a}$ are commuting fields while the others $c_{\alpha}^{i}$ and $\xi_{\alpha}^{\mu}$
      are anticommuting.

The action of the $\mathcal{N}=1$  supersymmetric generators
$\{P_{\mu},Q_{a}\}$ on these fields is given by
\begin{equation}
 [P_{\mu},X] = \partial_{\mu}X,         \nonumber\\
\end{equation}
\begin{equation}
[Q_{a},A^{i}_{\mu}] = -(\gamma_{\mu})_{ab}\lambda^{bi},\nonumber\\
\end{equation}
\begin{equation}
 [Q_{a},\lambda^{bi}] = -\frac{1}{2}(\sigma^{\mu\nu})_{a}^{b}F_{\mu\nu}^{i}+ \delta_{a}^{b}\Lambda^{i},\nonumber\\
\end{equation}
\begin{equation}
  [Q_{a},c_{\alpha}^{i}] = (\gamma^{\mu})_{ab}\chi_{\alpha}^{b}A_{\mu}^{i},\nonumber\\
\end{equation}
\begin{equation}
[Q_{a},\Lambda^{i}]=(\gamma^{\mu})_{ab}D_{\mu}\lambda^{bi},\nonumber\\
\end{equation}
\begin{equation}
[Q_{a},F_{\mu\nu}^{i}]=(\gamma_{\mu})_{ab}(D_{\nu}\lambda)^{bi}-(\gamma_{\nu})_{ab}(D_{\mu}\lambda)^{bi},\label{operators}
\end{equation}
where $X$ is any  field of the theory.

    It is worthwhile to mention that we are interested in our present investigation on the global
    supersymmetric transformations, so that the parameters of
    the $\mathcal{N}=1$ supersymmetric and translation groups must be space-time constant, i.e.
\begin{align}
   \partial_{\mu}\chi^{a}_{\alpha} = 0,\quad\quad\quad\quad
 \partial_{\mu}\xi_{\alpha}^{\nu} = 0.\label{parameters}
\end{align}
    Using the above identifications with (\ref{parameters}) and inserting the
    constraints (\ref{constraint1}) and (\ref{constraints2})
    into the Eqs. $(6)$  and (\ref{equationb2}), we obtain
\begin{equation}
\partial_{\alpha}\Phi^{i}_{\mu |}= (D_{\mu}c_{\alpha})^{i} - \xi_{\alpha}^{\nu}[P_{\nu},
A_{\mu}^{i}]- \chi_{\alpha}^{a}[Q_{a}, A_{\mu}^{i}],\nonumber\\
\end{equation}
\begin{equation}
\partial_{\alpha}\Phi^{i}_{\beta |}+\partial_{\beta}\Phi^{i}_{\alpha |}= -[c_{\alpha},c_{\beta}]^{i}
- \xi_{\alpha}^{\nu}[P_{\nu}, c_{\beta}^{i}]-
\chi_{\alpha}^{a}[Q_{a}, c_{\beta}^{i}]- \xi_{\beta}^{\nu}[P_{\nu},
c_{\alpha}^{i}]-\chi_{\beta}^{a}[Q_{a}, c_{\alpha}^{i}],\nonumber\\
\end{equation}
\begin{equation}
\partial_{\alpha}\Phi^{a}_{\beta |}+\partial_{\beta}\Phi^{a}_{\alpha |}=0,\nonumber\\
\end{equation}
\begin{equation}
\partial_{\alpha}\Phi^{\mu}_{\beta |}+\partial_{\beta}\Phi^{\mu}_{\alpha |}=-2\chi_{\alpha}^{a}(\gamma^{\mu})_{ab}\chi_{\beta}^{b},\nonumber\\
\end{equation}
\begin{equation}
\partial_{\alpha}\Omega^{i}_{\mu\nu |}= -[c_{\alpha}, F^{i}_{\mu\nu}] - \xi_{\alpha}^{\tau}[P_{\tau},
F_{\mu\nu}^{i}]- \chi_{\alpha}^{a}[Q_{a}, F_{\mu\nu}^{i}],\nonumber\\
\end{equation}
\begin{equation}
\partial_{\alpha}\Phi^{ai}_{|}= \frac{1}{4}(\gamma^{\mu})^{ab}[Q_{b},\partial_{\alpha}\Phi^{i}_{\mu |} ],\nonumber\\
\end{equation}
\begin{equation}
\partial_{\alpha}\Phi^{i}_{|}= -\frac{1}{4}(\delta^{a}_{b})[Q_{a},\partial_{\alpha}\Phi^{bi}_{ |} ].\label{commutators2}
\end{equation}
We also realize the usual identifications:
$Q_{\alpha}(X_{|})=\partial_{\alpha}X_{|}$, where X is any
superfield and $Q=Q_{1}$  $(\overline{Q}=Q_{2})$ is the BRST
(anti-BRST) operator. Inserting Eq. (\ref{operators}) into
(\ref{commutators2}), and evaluating these at $\theta^{\alpha}=0$,
we find the following BRST transformations
\begin{equation}
QA^{i}_{\mu}= (D_{\mu}c)^{i} - \xi^{\nu}\partial_{\nu}A^{i}_{\mu} + \chi\gamma_{\mu}\lambda^{i},\nonumber\\
\end{equation}
\begin{equation}
Q\lambda^{i}_{a}= -f^{i}_{jk} c^{j}\lambda^{k}_{a} -
\xi^{\mu}\partial_{\mu}\lambda^{i}_{a} + \chi_{a}\Lambda^{i}+
\frac{1}{2}(\chi\sigma^{\mu\nu})_{a}F^{i}_{\mu\nu},\nonumber\\
\end{equation}
\begin{equation}
Qc^{i}= -\frac{1}{2}f^{i}_{jk}c^{j} c^{k} -
\xi^{\mu}\partial_{\mu}c^{i} +
\chi\gamma^{u}\overline{\chi}A^{i}_{\mu},\nonumber\\
\end{equation}
\begin{equation}
Q\xi^{\mu}= -\chi\gamma^{u}\overline{\chi},\nonumber\\
\end{equation}
\begin{equation}
Q\Lambda^{i}= -f^{i}_{jk} c^{k}\Lambda^{j} -
\xi^{\rho}\partial_{\rho}\Lambda^{i} -\chi\gamma^{\mu}D_{\mu}\lambda^{i},\nonumber\\
\end{equation}
\begin{equation}
QF_{\mu\nu}^{i}= -f^{i}_{jk} c^{k}F^{j}_{\mu\nu} -
\xi^{\rho}\partial_{\rho}F_{\mu\nu}^{i} -\chi^{a}\{(\gamma_{\mu})_{ab}(D_{\nu}\lambda)^{bi}-
(\gamma_{\nu})_{ab}(D_{\mu}\lambda)^{bi}\},\nonumber\\
\end{equation}
\begin{equation}
Q\chi^{a}=0, \quad Q\overline{c}^{i}= B^{i},\quad QB^{i}= 0,\quad
Q\overline{\xi}^{\mu}=E^{\mu},\quad QE^{\mu}= 0,\quad
Q\overline{\chi}^{a}= G^{a},\quad Q G^{a}= 0,\label{transformations
BRST}
\end{equation}
and also the anti-BRST  transformations, which can be derived from
(\ref{transformations BRST}) by the following mirror symmetry of the
ghost numbers given by : $X\rightarrow X$  if $X=
A^{i}_{\mu},\lambda^{i}_{a}, \Lambda^{i}$; $X\rightarrow
\overline{X}$ if $X= Q, c^{i}, B^{i}, \xi^{\mu},
 E^{\mu}, \chi^{a}, G^{a}$ and  $\overline{\overline{X}}=X$ where
\begin{equation}
B^{i}+ \overline{B}^{i}= -f^{i}_{jk} c^{k}\overline{c}^{j}
-\overline{\xi}^{\nu}\partial_{\nu}c^{i}-
\xi^{\mu}\partial_{\mu}\overline{c}^{i} -
\chi\gamma^{\mu}\overline{\chi}A^{i}_{\mu}-
\overline{\chi}\gamma^{\mu}\chi A^{i}_{\mu},\label{mirror sym1}
\end{equation}
\begin{equation}
E^{\mu}+ \overline{E}^{\mu}= -2 \chi\gamma^{\mu}\overline{\chi}
,\quad\quad\quad\quad\quad\quad G^{a}+ \overline{G}^{a}=
0.\label{mirror sym2}
\end{equation}
Let us note that the introduction of an auxiliary real field
$\Lambda^{i}$ besides the fields present in quantized
$\mathcal{N}=1$ super Yang-Mills theory in four-dimensions,
guarantees automatically the off-shell nilpotency of the $
\{Q,\overline{Q}\}$ algebra and make easier then, as we will see in
the next section, the gauge-fixing process.
\section {Quantum action}
 \bigskip In the present section, we show how to construct in the context of
our procedure a BRST-invariant quantum action for $\mathcal{N}=1$
super Yang-Mills theory as the lowest component of a quantum
superaction. To this purpose,  we choose the following gauge-fixing
superaction
\begin{equation}
S_{sgf} = \int d^{4}x L_{sgf},\nonumber\\
\end{equation}
\begin{equation}
L_{sgf} = (\partial_{1}\Phi_{2})(\partial^{\mu}\Phi_{\mu})+
(\partial^{\mu}\Phi_{2})(\partial_{1}\Phi_{\mu})+
    (\partial_{1}\Phi_{2})(\partial_{1}\Phi_{2}).\label{Lagrangien sgf}
\end{equation}
Let us recall that similar gauge-fixing superaction was used in
Refs. \cite{tah1,tah2,meziane}. We note first that in the case of
Yang-Mills theory the superaction involves a Lorentz gauge
\cite{tahiri11} given by
\begin{equation}
\partial_{\mu}\Phi^{\mu}_{|}= 0. \label{lorentz gauge}
\end{equation}
In the case of super Yang-Mills theory we shall choose a
supersymmetric gauge-fixing which is the extension of the Lorentz
gauge. This gauge fixing can be obtained from (\ref{lorentz gauge})
by using the following substitution
\begin{equation}
\Phi_{\mu}\rightarrow  \widetilde{\Phi_{\mu}}=\Phi_{\mu} +
[\partial_{\mu}\Phi^{a},Q_{a}].
\end{equation}
Now, it is easy to see that the gauge-fixing superaction
(\ref{Lagrangien sgf}) can be put in the following form
\begin{equation}
S_{sgf} = \int d^{4}x[
(\partial_{1}\Phi_{2})(\partial^{\mu}\widetilde{\Phi_{\mu}})+
(\partial^{\mu}\Phi_{2})(\partial_{1}\widetilde{\Phi_{\mu}})].\label{gaugefixing}
\end{equation}
To determine the gauge-fixing action $S_{gf}$ as the lowest
component of the gauge-fixing superaction $S_{gf}=S_{sgf|}$, we
impose the following rules
\begin{equation}
Tr(I^{m}I_{n})  = \delta_{n}^{m},\nonumber\\
\end{equation}
\begin{equation}
Tr([Q_{a},Q_{b}])  = 2(\gamma^{\mu})_{ab}\partial_{\mu},\nonumber\\
\end{equation}
\begin{equation}
Tr(P^{2})  = 0.\label{rules}
\end{equation}
These rules permit us to compute the trace of each term in
(\ref{gaugefixing}) . Indeed, from (\ref{rules}) it is easy to put
the gauge-fixing action $S_{^{gf}}$ in the form

\begin{equation}
S_{^{gf}}=S_{^{sgf}|} = \int d^{4}x [B\partial^{\mu}A_{\mu} +
2b^{*}_{j} G(\gamma^{\mu}\partial_{\mu}\Box\lambda ^{j})\nonumber\\
\end{equation}
\begin{equation}
+(\partial^{\mu}\overline{c})\{D_{\mu}c + \xi^{\nu}\partial_{\nu}
A_{\mu}+ \chi\gamma_{\mu}\lambda\}\nonumber\\
\end{equation}
\begin{equation}
-2b^{*}_{j}(\partial^{\mu}\overline{\chi})\gamma^{\nu}\partial_{\nu}\partial_{\mu}\{
f^{j}_{ik} \lambda^{i}c^{k} + \xi^{\tau}\partial_{\tau}\lambda^{j}
-\frac{1}{2}\chi\sigma^{\tau\nu}F^{j}_{\tau\nu} -\chi\Lambda^{j}\}].
\end{equation}

On the other hand, the presence of the extrafield breaks the
invariance of the classical action ({\ref{classical
 action1}}). In fact, the only terms which may contribute to the $Q$-variation of the classical
action $S_{0}$ are those containing the extrafield $\Lambda ^{i}$.
This follows from the fact that the BRST transformations up to terms
$\Lambda ^{i}$ represent the $\mathcal{N}=1$ super Yang-Mills
transformations expressed \`{a} la BRST. A simple calculation with
the help of the BRST transformations (\ref{transformations BRST})
leads to
\begin{equation}
QS_{0}=\chi^{a}\Lambda^{i}(\gamma^{\mu})_{ab}(D_{\mu}\lambda^{b})_{i}.\label{variation
actionclassic}
\end{equation}
Thus the classical action $S_{0}$ is not BRST-invariant, and in
order to find the BRST-invariant extension $S_{inv}$ of the
classical action, we shall add to $S_{0}$ a term $\widetilde{S_{0}}$
so that
\begin{equation}
 Q(S_{0} + \widetilde{S_{0}})=0.\label{variation
actionclassic2}
\end{equation}
We remark that  $\widetilde{S_{0}}$ is the part of the extended
classical action related  to the  auxiliary field $\Lambda ^{i}$ and
 is given by
\begin{equation}
\widetilde{S_{0}}= -\frac{1}{2}\Lambda^{i}\Lambda_{i}.
 \end{equation}
Then, it is quite easy to show that $Q(S_{0})=-Q(\widetilde{S_{0}})$
by a direct calculation with the help of the transformations
(\ref{transformations BRST}).

Having found the BRST-invariant extended action $S_{inv}$ we now
write the full off-shell BRST- invariant quantum action $S_{q}$ by
adding to the $Q$-invariant action, $S_{inv}=S_{0} +
\widetilde{S_{0}} $,  the $Q$-invariant gauge-fixing action $S_{gf}$
\begin{equation}
S_{q}= S_{0} + \widetilde{S_{0}}+S_{gf}.\label{quantum action}
 \end{equation}

 It is worth nothing that the quantum action (\ref{quantum action})
 allows us to see that the auxiliary field $\Lambda ^{i}$ does not
propagate, as its equation of motion is a constraint
\begin{equation}
\frac{\delta S_{gf}}{\delta \Lambda ^{i}}= -\Lambda_{i}+
2b^{*}_{i}(\widetilde{\chi}\gamma^{\mu}\partial_{\mu}\Box\chi)=
0.\label{algebrique constraint}
\end{equation}
Thus the essential role of the nondynamical auxiliary field $\Lambda
^{i}$ is to close the BRST and anti-BRST algebra off-shell.

 The elimination of the auxiliary field $\Lambda ^{i}$ by means of its equation
 of motion (\ref{algebrique constraint}) leads to the same gauge-fixed theory with on-shell
 nilpotent BRST transformations obtained in the context of BV formalism
  \cite{Baulieu} as well as in the framework of the superfibre bundle approach \cite{tah1}.

 Moreover, in our formalism we have also introduced an anti-BRST operator $\overline{Q}$ and it
 is important to realize that both the BRST symmetry and anti-BRST symmetry can be
 taken into account on an equal footing. To this end, we simply use the fact that
  there is a complete duality, with respect to the mirror symmetry of the ghost number, between the
  $Q $ and $\overline{Q}-$ transformations. So, the $\overline{Q}$-variation of the
classical action $S_{0}$ is given by
\begin{equation}
\overline{Q}S_{0} =\overline{\chi}^{a}\Lambda^{i}(
\gamma^{\mu})_{ab}(D_{\mu}\lambda^{b})_{i}  .
  \end{equation}
  Using however the $\overline{Q}$-transformations of the auxiliary field (see Eqs.
 (\ref{transformations BRST}) with the mirror symmetry), we obtain that the $Q$-invariant action
  $S_{inv}=S_{0}+S_{0} $  is also $\overline{Q}$-invariant.
  Furthermore, the $Q$-gauge-fixing action can be also written as in Yang-Mills theories in $\overline{Q}$-form.
   Therefore the full off-shell BRST-invariant quantum action $S_{q}=S_{0}+ \widetilde{S_{0}}+S_{gf}$ is
  also an off-shell anti-BRST-invariant quantum action.

\section{ Conclusion}
\bigskip In the present paper we have developed a BRST superspace formalism
in order to perform the quantization of the four dimensional
$\mathcal{N}=1$ supersymmetric Yang-Mills theory as model where the
classical gauge algebra is not closed. In this geometrical
framework, our construction is entirely based on the possibility of
introducing \emph{ab initio} a set of fields through a super
Yang-Mills connection. The latter represents the gauge fields and
their associated ghost and antighost fields occurring in such
theory, whereas the extrafield coming from  the superconnection via
the  supersymmetric transformations is required to achieve the
off-shell nilpotency of the BRST and anti-BRST operators. Let us
note that for a local symmetry  the minimal set of auxiliary fields
is introduced by the supercurvature while for our case of global
symmetry the auxiliary real field  is introduced via the
superYang-Mills connection.

Furthermore, we have performed a direct construction of the BRST
invariant gauge fixed  action for $\mathcal{N}=1$ , $4 D$
supersymmetric Yang-Mills theory in analogy with what it is realized
in BF theories \cite{tahiri2} and simple supergravity
\cite{tahiri22}.  The obtained quantum action allows us to see that
the extrafield enjoy the auxiliary freedom.  The elimination of this
auxiliary field using its equation of motion permits us to recover
the standard quantum action with on-shell nilpotent BRST symmetry
\cite{Baulieu}.  By using the mirror symmetry between the BRST and
anti-BRST transformations, we can see that the BRST invariant
extended classical action is also anti-BRST invariant. Therefore the
full quantum action is BRST and anti-BRST invariant, since the
gauge-fixing action can be written as in the Yang-Mills case in BRST
as well as anti-BRST exact form, due to the off-shell nilpotency of
the BRST-anti-BRST algebra.

Finally, we should mention that the BRST superspace formalism
represents the natural arena where the fields and their off-shell
nilpotent BRST and anti-BRST transformations for gauge theories can
be found.  This is not only the case of Yang-Mills type theories,
arbitrary gauge theories may be also treated in this framework.
Indeed, such formalism was  applied to several interesting theories
with local symmetry such non-Abelian BF theory \cite{tahiri2} and
simple supergravity \cite{tahiri22}. In the present work this
formalism has been applied successfully to the theory with a global
supersymmetry. The off-shell nilpotency is naturally implemented
through the introduction of auxiliary field required for the
consistency of the BRST superspace geometry. Thus, it would be a
very nice endeavor to use this basic idea to study the structure of
auxiliary field in other gauge theories.

\section{Acknowledgments}
A. Meziane would like to thank  Prof. H. Nicolai, director at Max
Planck Institute for Gravitational Physics, Albert Einstein
Institute, Germany,  for his warm hospitality and illuminating
discussions. The DAAD is gratefully acknowledged too for his
financial support.

\section{REFERENCES}

\bibliographystyle{amsplain}

\providecommand{\bysame}{\leavevmode\hbox to3em{\hrulefill}\thinspace}
\providecommand{\MR}{\relax\ifhmode\unskip\space\fi MR }
\providecommand{\MRhref}[2]{%
  \href{http://www.ams.org/mathscinet-getitem?mr=#1}{#2}
}
\providecommand{\href}[2]{#2}
\begin{thebibliography}{}

\end{thebibliography}


\begin{thebibliography}{10}
\bibitem{Hen}
M. Henneaux and C. Teitelboim, \textit{Quantization of Gauge
Systems}, Princeton University Press, Princeton, New Jersey, (1992).
\bibitem{maison}
P. Breitenlohner and D. Maison, \emph{ Renormalization of
supersymmetric Yang-Mills theories,} in Cambridge Proceedings
(1985), \emph{Supersymmetry and its applications,} p. 309;  and
\emph{$\mathcal{N}=2$ Supersymmetric Yang-Mills theories in the
Wess-Zumino gauge, in Renormalization of quantum field theories with
nonlinear field transformations}, Proceedings, workshop, Tegernsee
(1987), by P. Breitenlohner, (Ed.), D. Maison (Ed.),  K.
Sibold,(Ed.) Springer (1988) p. 64.
\bibitem{magiorenico}
N. Maggiore, \emph{Int. J. Mod. Phys.} {\bf A 10}, (1995) 3937; {\bf
A 10}, (1995) 3781; N. Maggiore, O. Piguet and S. Wolf, \emph{Nucl.
Phys.} {\bf B 458}, (1996) 403; {\bf B 476}, (1996) 329.
\bibitem{dixon1}
J. A. Dixon,  \emph{Class. Quant. Grav.} {\bf 7}, (1990) 1511.
\bibitem{white}
P. L. White, \emph{Class. Quant. Grav.} {\bf 9}, (1992) 413.
 \bibitem{magiore1}
 N. Maggiore and M. Schaden, \emph{Phys. Rev.} {\bf D 50}, (1994) 6616.
 \bibitem{dixon2}
 J. A. Dixon, \emph{Commun. Math. Phys.} {\bf 140}, (1991) 169.
\bibitem{Van-Nieuwen}
P. van Nieuwenhuizen, \emph{Phys. Rep.} {\bf 68}, (1981) 189 .
\bibitem{bat22}
 I. A. Batalin and G. A. Vilkovisky, \emph{Phys. Lett.} {\bf B 102}, (1981) 27; \emph{Phys. Rev. }{\bf D 28}, (1983)2567.
 \bibitem{Baulieu}
L. Baulieu, M. Bellon, S. Ouvry and J. C. Wallet, \emph{Phys. Lett.}
{\bf B 252}, (1990) 387.
\bibitem{Galperin}
A. Galperin, E. Ivanov, S. Kalitzin, V. Ogievetsky and E. Sokatchev,
\emph{Class. Quant. Grav.} {\bf 1}, (1984) 469.
\bibitem{Galperin2}
 A. Galperin, E. Ivanov, S. Kalitzin, V. Ogievetsky and E.
Sokatchev, \textit{Harmonic Superspace}, Cambridge University Press,
(2001).
\bibitem{Bonora}
L. Bonora and M. Tonin, \emph{Phys. Lett.} {\bf B 98},  (1981) 48.
\bibitem{Malik}
R.P. Malik, \emph{ J. Phys. Math.} {\bf 3}, P110503 (2011).
\bibitem{tahiri2}
M. Tahiri, \emph{Phys. Lett.} {\bf B 325}, (1994)  71;  \emph{Int.
J. Mod. Phys.} {\bf A 12}, (1997)  3153.
\bibitem{tahiri22}
M. Tahiri, \emph{Phys. Lett.} {\bf B 403},  (1997) 273.
\bibitem{Wess}
J. Wess and J. Bagger, \textit{Supersymmetry and Supergravity},
Princeton University Press, Princeton, New Jersey, (1983).
\bibitem{WEinberg}
S. Weinberg, \textit{The Quantum Theory of Fields, Vol.3:
Supersymmetry}, Cambridge University Press, New York, (2000).
\bibitem{tah1}
H. Loumi-Fergane and M. Tahiri, \emph{Class. Quant. Grav.}{\bf 13},
 (1996) 865.
\bibitem{tah2}
A. Aidaoui and M. Tahiri, \emph{Class. Quant. Grav.} {\bf 14},
(1997) 1587.
\bibitem{meziane}
A. Meziane and M. Tahiri, \emph{Phys. Rev.} {\bf D 71}, (2005)
104033.
 \bibitem{tahiri11}
 H. Loumi and M. Tahiri, \emph{Rep. Math. Phys. } {\bf 33}, (1993) 367 .

\end{thebibliography}

\end{document}